\documentclass[aps,prl,reprint,superscriptaddress,showpacs]{revtex4-1}
\usepackage{graphicx}
\usepackage{amsmath}
\usepackage{amsthm}
\usepackage{xcolor}
\usepackage{txfonts}
\usepackage{colortbl}
\usepackage[colorlinks=true,citecolor=blue,linkcolor=violet]{hyperref}

\newcommand{\Rayleigh}{\text{Ra}}
\newcommand{\Prandtl}{\text{Pr}}
\newcommand{\Ekman}{E}

\begin{document}
\graphicspath{{./figures/}}
\title{Approaching the Asymptotic Regime of Rapidly Rotating Convection: \\Boundary Layers vs Interior Dynamics}
\author{S. Stellmach}
\email[]{stephan.stellmach@uni-muenster.de}
\affiliation{Institut f\"ur Geophysik, Westf\"alische Wilhelms-Universit\"at M\"unster, Germany,}
\author{M. Lischper}
\affiliation{Institut f\"ur Geophysik, Westf\"alische Wilhelms-Universit\"at M\"unster, Germany,}
\author{K. Julien}
\affiliation{Department of Applied Mathematics, University of Colorado Boulder, Boulder, CO 80309, USA}
\author{G. Vasil}
\affiliation{School of Mathematics and Statistics, University of Sydney, Australia}
\author{J.S. Cheng}
\affiliation{Department of Earth, Planetary and Space Sciences, University of California, Los Angeles, CA. 90095-1567}
\author{A. Ribeiro}
\affiliation{Department of Earth, Planetary and Space Sciences, University of California, Los Angeles, CA. 90095-1567}
\author{E. M. King}
\affiliation{Miller Institute and Dept. of Earth and Planetary Science, Berkeley, CA, USA}
\author{J. M. Aurnou}
\affiliation{Department of Earth, Planetary and Space Sciences, University of California, Los Angeles, CA. 90095-1567}
\date{\today}
\begin{abstract}
Rapidly rotating Rayleigh-B\'enard convection is studied by combining results from direct numerical simulations (DNS), laboratory experiments and asymptotic modeling. The asymptotic theory is shown to provide a good description of the bulk dynamics at low, but finite Rossby number. However, large deviations from the asymptotically predicted heat transfer scaling are found, with laboratory experiments and DNS consistently yielding much larger Nusselt numbers than expected. These deviations are traced down to dynamically active Ekman boundary layers, which are shown to play an integral part in controlling heat transfer even for Ekman numbers as small as $10^{-7}$. By adding an analytical parameterization of the Ekman transport to simulations using stress-free boundary conditions, we demonstrate that the heat transfer jumps from values broadly compatible with the asymptotic theory to states of strongly increased heat transfer, in good quantitative agreement with no-slip DNS and compatible with the experimental data. Finally, similarly to non-rotating convection, we find no single scaling behavior, but instead that multiple well-defined \textcolor{black}{dynamical} regimes exist in rapidly-rotating convection systems.
\end{abstract}
\pacs{47.20.Bp, 47.32.Ef, 47.55.pb, 47.27.-i}
\maketitle

Rapidly rotating thermal convection is ubiquitous in nature. It occurs in the ocean, in the liquid metal cores of terrestrial planets, in gas giants and in rapidly rotating stars. All these systems are highly turbulent, but at the same time Coriolis forces chiefly control their dynamics. It is this dominating role of Coriolis forces which gives convection in many large-scale natural systems its distinctive character. 

Different from non-rotating convection, where large regions of the parameter space have been explored extensively over the last decades \cite{ahlers2009heat}, both experiments and direct numerical simulations (DNS) face serious difficulties in entering the turbulent, but rotationally constrained regime. While experiments easily reach high levels of turbulence, they struggle to ensure that Coriolis forces remain dominant in the force balance  \citep{king2012heat,ecke2014heat,cheng2014laboratory}. 
Numerical simulations suffer from the enormous range of spatial and temporal scales that need to be resolved. As a consequence, the available data is scarce, and the scaling laws that are needed for quantifying the effects of convection in large-scale natural systems remain poorly constrained. 

The canonical framework to study rotating convection is the rotating Rayleigh-B\'enard system. 
A plane fluid layer of depth $H$, destabilized by a constant temperature difference $\Delta T$ between the boundaries, rotates about a vertical axis with angular velocity $\Omega$. Within the Boussinesq approximation, three non-dimensional parameters control the system behavior. The Rayleigh number $Ra = g \alpha \Delta T H^3 / \kappa \nu$, where $g$ denotes gravitational acceleration,  $\alpha$ the thermal expansion coefficient, $\nu$ the kinematic viscosity and $\kappa$ the thermal diffusivity, measures the forcing strength. The Ekman number $E=\nu / 2 \Omega H^2$ is defined as the ratio of the rotational time scale to the viscous diffusion time scale. Finally, the Prandtl number $Pr = \nu / \kappa$ signifies the efficiency of viscous relative to thermal diffusion. A combination of these parameters, the convective Rossby number $Ro_c = \sqrt{\Rayleigh / \Prandtl} \, \Ekman$ is often used as a proxy for the importance of rotation relative to the thermal forcing. 

Rapidly rotating convection is characterized by $Ro_c \ll 1$, with the asymptotic limit $Ro_c \rightarrow 0$ representing an important limiting case. Previous theoretical work \cite{julien1998new,sprague2006numerical} has shown that the governing equations can be simplified substantially in this limit. The resulting reduced set of equations, essentially a non-hydrostatic quasi-geostrophic model, is expected to hold for $Ra \le O(E^{-5/3})$ and $\Pr \ge O(E^{1/4})$ \cite{julien2012statistical}. The Ekman layer is assumed to become passive at small $E$ \cite{niiler1965influence,julien1998strongly}, and its $O(E^{1/2})$ vertical transport is assumed to be negligible for the dynamics in this regime \citep{dawes2001rapidly,sprague2006numerical,julien2012heat}. Numerical simulations using the reduced equations have revealed a rich dynamical behavior \cite{sprague2006numerical,julien2012statistical,julien2012heat}, characterized by the existence of several distinct flow regimes, each associated with different heat transfer properties. 

Previous experiments \cite{rossby1969study,liu1997heat,zhong2010heat,king2009boundary} and DNS \cite{schmitz2009heat,king2009boundary}  barely reach into the rapidly rotating regime, and their relation to the asymptotic model is thus unclear.  Here, we attempt to establish such a relation by pushing direct numerical simulations and laboratory experiments further into the rapidly rotating regime, where a comparison to asymptotic models becomes meaningful. A central result of our study is that the behavior in the {\it bulk} is well described by the asymptotic model, while, contrary to common expectations, the {\it viscous boundary layer dynamics} largely controls the heat transfer scaling whenever the fluid layer is confined between no-slip boundaries. In fact, even for the lowest Ekman numbers that can be reached in laboratory experiments and DNS today, viscous boundary layers are shown to massivley boost the heat transfer  in the low Rossby number regime, leading to a considerable increase of the exponent $\alpha$ in power laws of the form $Nu=f(Pr,Ek) \, Ra^{\alpha}$, where the Nusselt number $Nu$ is defined as the total heat flow normalized by the purely conductive value.
\begin{figure}
\centering
\centering
\includegraphics[width=0.9\linewidth]{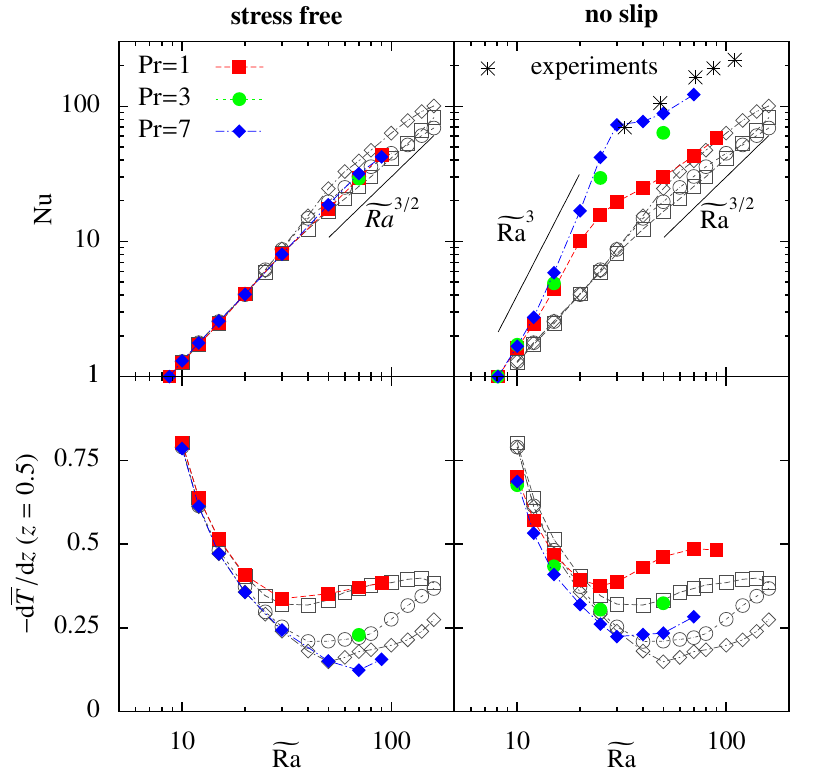}
\caption{Nusselt number and interior temperature gradient at the midplane as a function of $\widetilde{\text{Ra}}$. The filled symbols show DNS results at $E=10^{-7}$, while the open symbols are asymptotic predictions \cite{julien2012heat}. The stars shows laboratory values obtained in a cylindical rotating water tank with $5.79 \le \Prandtl \le 6.2$ and $10^{-7} \le \Ekman \le 1.1 \times 10^{-7}$.}
\label{fig:Nu_Ra}
\end{figure}

To illustrate this, figure \ref{fig:Nu_Ra} shows the Nusselt numbers found in DNS and in experiments at $E\approx 10^{-7}$.   In order to compare our results with the asymptotic predictions, we use $\widetilde{Ra} = Ra E^{4/3}$ as a measure for the super-criticality of the system \cite{sprague2006numerical}. The DNS employ both no-slip and stress-free boundary conditions, and are carried out in a horizontally periodic box with an aspect ratio $\Gamma = 10 \, l_\perp H^{-1}$, where $l_\perp$ is the critical wavelength of the marginally unstable mode for the stress-free case. Resolutions up to $576 \times 576 \times 513$ grid points are used to fully resolve the flow down to the dissipation scale. Care has also been taken to resolve the thin Ekman layers, characterized by an $O(\Ekman^{1/2} \,H) $ thickness, with a minimum of ten grid points. Experiments, made using water with $Pr \approx 7$ as the working fluid, are carried out in axially-aligned cylindrical containers with diameters of 18.73 cm and heights of 40, 80 and 160 cm.  A detailed description of the numerical and experimental techniques are, respectively, given in \cite{stellmach2008efficient} and \cite{king2012heat,cheng2014laboratory}.  

For stress-free boundary conditions and moderate $\widetilde{Ra}$, fig. \ref{fig:Nu_Ra} reveals an excellent agreement between DNS and asymptotic results, which becomes poorer for $\widetilde{Ra} > 50$. The results are broadly compatible with $\alpha = 3/2$ at high $\widetilde{Ra}$, as expected for diffusion independent turbulent heat transfer \cite{julien2012heat}. The interior temperature gradient found in the DNS follows the asymptotic results closely over the entire range of $\widetilde{Ra}$ considered in this study. In contrast, DNS and asymptotic results diverge strongly for no-slip boundary conditions. From onset on, the DNS reveals a much steeper heat transfer scaling, broadly compatible with the $\alpha ~ \sim 3$ scaling proposed in \cite{king2012heat}, which flattens out as soon as the interior temperature gradient reaches its minimum. At this point, the different scaling behavior has resulted in a Nusselt number that exceeds the corresponding values obtained for stress-free boundaries by more than $800\%$ for $Pr=7$. Data from laboratory experiments, denoted by stars in fig. \ref{fig:Nu_Ra}, is in line with these numerical findings. 

 \begin{figure}
\centering
\includegraphics[width=0.85\linewidth]{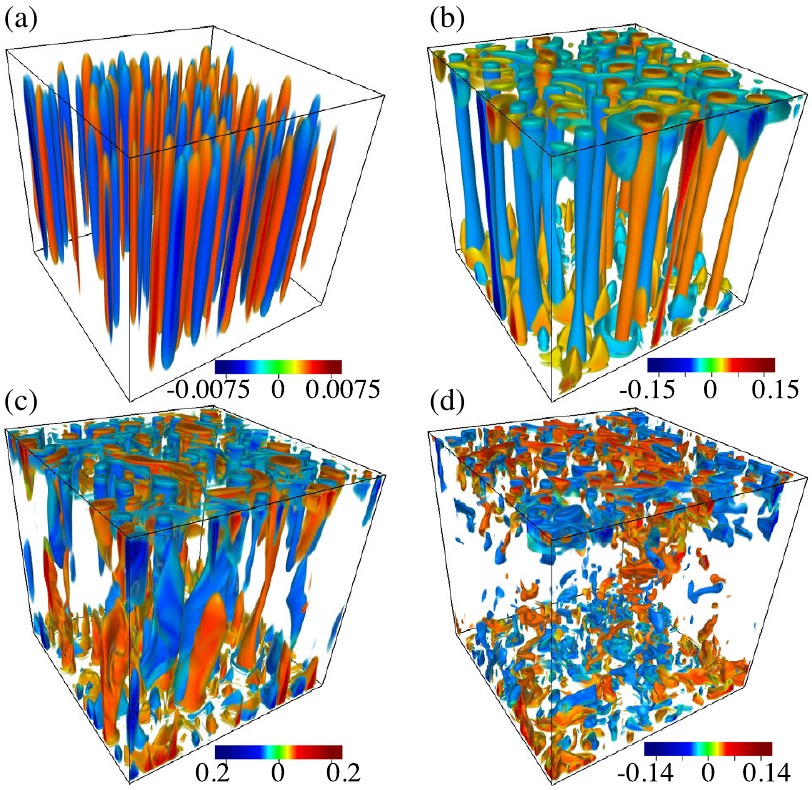}
\caption{Thermal anomaly  $\theta = T - \overline{T}$ in DNS at $E=10^{-7}$ with no-slip boundary conditions. 
(a) Cellular regime, $\widetilde{\text{Ra}}=10,Pr=1$, (b) convective Taylor columns, $\widetilde{\text{Ra}}=25, \text{Pr}=7$, (c) plumes, $\widetilde{\text{Ra}}=70, \text{Pr}=7$ and (d) geostrophic turbulence,  $\widetilde{\text{Ra}}=90, \text{Pr}=1$. For better visibility, the domain has been stretched horizontally by a factor of $4.5$. $\theta$ is scaled with $\Delta T$ in all cases.}
\label{fig:snapshots}
\end{figure}
The excellent agreement found between DNS with stress-free boundaries and the asymptotic results suggests that the {\it bulk dynamics} is well captured by the asymptotic model. This conjecture is supported by the fact that  the reduced model also provides a qualitatively correct picture of the dominant flow regimes in the bulk of the convective layer for no-slip DNS. For small $\widetilde{\Rayleigh}$, a cellular regime is found (fig. \ref{fig:snapshots}a), which gives way to a regime characterized by so-called convective Taylor columns (CTCs) \cite{grooms2010model} as $\widetilde{\Rayleigh}$ is increased (fig. \ref{fig:snapshots}b). The up- and downwellings form intense columnar structures, which are less densely packed than in the cellular regime. Their intense vortex cores are shielded by weaker sheets of opposite vorticity and temperature anomaly, preventing them from being destroyed by vortex-vortex interactions. The interior temperature gradient decreases with $\widetilde{\Rayleigh}$ through the celluar and CTC regime. Its downward trend is finally broken when the CTCs begin to loose their shields, which exposes the vortex cores. Violent vortical interactions then destroy the vertically coherent columns and lead to the formation of plume-like structures (fig. \ref{fig:snapshots}c), which is accompanied by an increasing interior temperature gradient.  For $Pr=1$, the system directly evolves from the cellular regime to this plume state, without forming CTCs. A further increase of $\widetilde{\Rayleigh}$ then leads to a complete breakdown of vertical coherence for $Pr=1$ (fig. \ref{fig:snapshots}d), the interior temperature gradient saturates and the flow enters a regime called geostrophic turbulence (GT). Similar regimes are also observed in the stress-free case, with the regime boundaries agreeing well with the asymptotic model even quantitatively.\\
\begin{figure}
\centering
\includegraphics[width=0.85\linewidth]{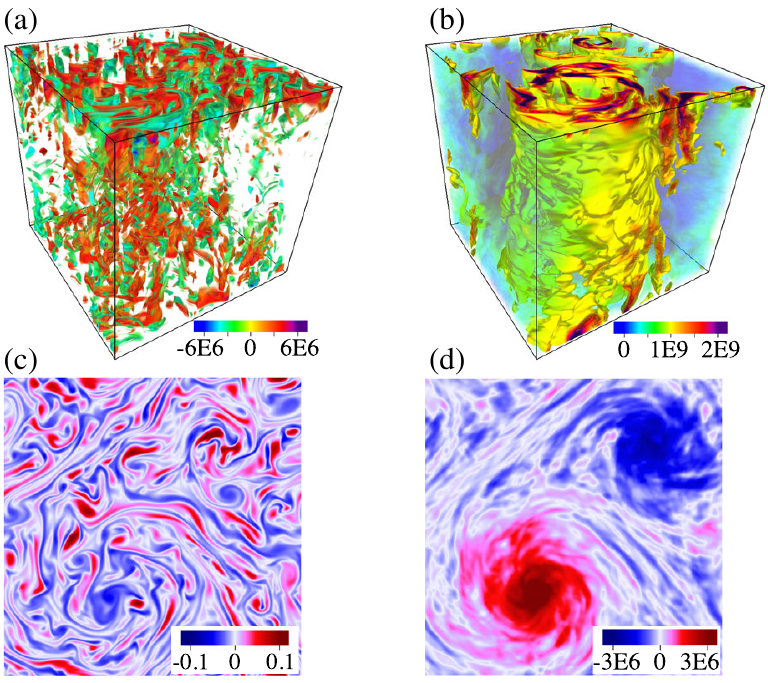}
\caption{Formation of large-scale, barotropic vortices of both signs of vorticity in the GT regime for stress-free boundaries at $E=10^{-7},\widetilde{\text{Ra}}=90, \text{Pr}=1$. Shown is (a) vertical vorticity , (b) horizontal kinetic energy, (c) temperature anomaly at $z=0.99$ and (d) vertically averaged vertical vorticity. \textcolor{black}{Units are $\kappa/H^2$, $\kappa^2/H^2$, $\Delta T$ and $\kappa/H^2$, respectively.}}
\label{fig:barotropic mode}
\end{figure}
An interesting feature of the GT regime is the formation of large-scale barotropic vortices driven by an upscale transport of kinetic energy \cite{julien2012statistical,rubio2014upscale}. The effect is most clearly observed in the stress-free case
(fig \ref{fig:barotropic mode}). The formation of large scale vortices in rotating convection has recently been observed in simulations at larger $E$  \cite{guervilly2014large,favier2014inverse}, where the large-scale energy accumulates predominantly in cyclonic structures. Such symmetry-breaking is predicted to be absent in the asymptotic small Rossby number case  \cite{rubio2014upscale}. Indeed, our DNS reveals the generation of both\textcolor{black}{, strong} cyclonic and anti-cyclonic vortices, which shows that symmetry \textcolor{black}{slowly} tends to be restored with decreasing Rossby number. \textcolor{black}{The formation of large scale coherent vortices is inhibited in DNS that employ no-slip boundaries, similar to the results of \cite{ostilla_monico20014geostrophic}}.
\begin{figure}
\centering
\includegraphics[width=\linewidth]{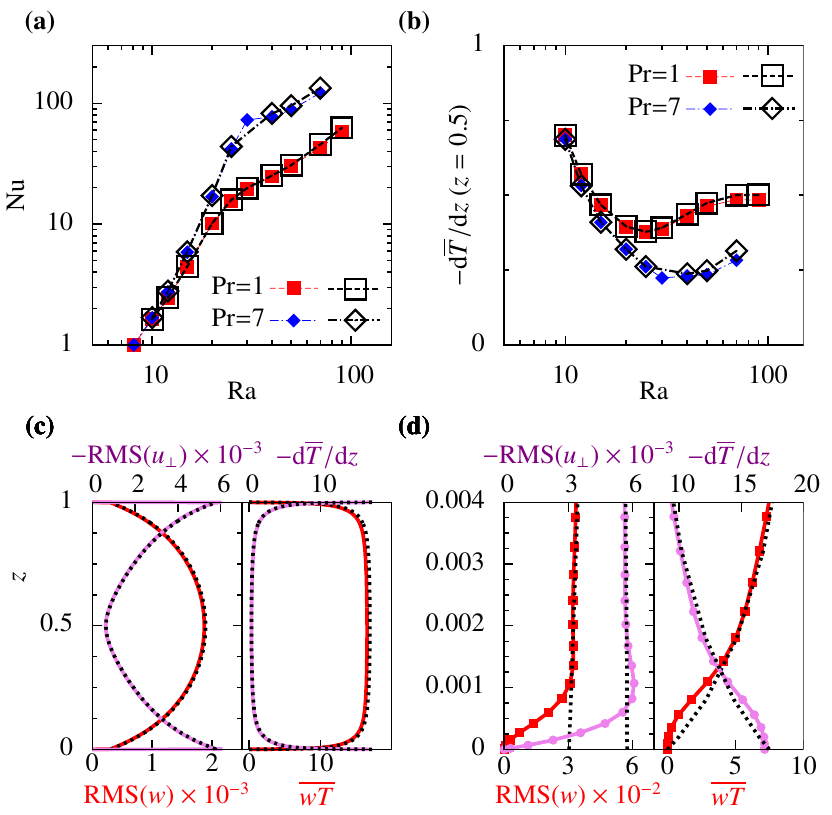}
\caption{Results obtained with parameterized Ekman pumping. Filled symbols \textcolor{black}{denote no-slip cases}, while open symbols show results obtained \textcolor{black}{by applying} the pumping boundary conditions (\ref{eq:BCs_pumping}) \textcolor{black}{to the full equations}. 
Vertical profiles for \textcolor{black}{$E=10^{-7}$}, $Pr=7, \widetilde{Ra}=20$ are shown in (c), with a boundary layer blowup  in (d) for no-slip (solid) and Ekman pumping (dashed) boundary conditions. ($u_\perp$: horizontal velocity, $w$: vertical velocity, $T$:temperature, given in units of $\kappa/H$ and $\Delta T$, respectively).}
\label{fig:EkmanPumping}
\end{figure}

Our results thus suggest that at $E=10^{-7}$, the bulk dynamics shows clear signs of asymptotic behavior, while viscous boundary layers strongly increase the heat transfer efficiency, leading to a different scaling behavior than predicted asymptotically. The mechanical boundary conditions thus exert a controlling influence on the heat transfer scaling, a result which is unexpected for the low Ekman number considered here. Studies of the linear problem \cite{niiler1965influence} suggest that the onset of convection becomes largely independent of the mechanical boundary conditions for $\Ekman \le 10^{-6}$. While non-linear contributions from Ekman pumping are known to increase the heat transport at moderate Ekman numbers \cite{kunnen2006heat}, it is generally expected that this effect becomes small in the low $E$ regime, as both the Ekman layer thickness and the associated secondary flow should decrease with $O(E^{1/2})$, such that the Ekman layer finally becomes passive \citep{dawes2001rapidly,sprague2006numerical,julien2012heat}. It is remarkable that even at $E=10^{-7}$, where the Ekman layer covers only about $0.1 \%$ of the layer depth, an increase by  almost an order of magnitude in the heat tranfer is observed. 


The effects of no-slip boundaries can be modeled using boundary conditions that parameterize fluid pumping into and out of the linear Ekman layers \cite{greenspan1990theory}. Pumping is added to otherwise stress-free boundary conditions by enforcing 
\begin{equation}
\frac{\partial u_x}{\partial z} =  \frac{\partial u_y}{\partial z} = 0 , \quad u_z = \pm \left(\frac{E}{2} \right)^{1/2} \omega_z H \quad \text{at} \quad z=0,H
\label{eq:BCs_pumping}
\end{equation}
\textcolor{black}{in DNS}, where $\omega_z$ denotes vertical vorticity. The plus sign applies to the lower and the minus sign to the upper boundary, respectively. Note that this choice isolates the Ekman pumping effects, while neglecting the viscous dissipation in the boundary layer that plays a central role in theories of non-rotating convection \cite{grossmann2000scaling}. Figure \ref{fig:EkmanPumping} shows the results obtained. Both the heat transfer and the interior temperature gradients found in the DNS are well captured by the pumping parameterization (figure \ref{fig:EkmanPumping} a,b). Panels (c) and (d) show profiles of the rms horizontal and vertical velocity and of conductive and advective heat transfer for the special case $Pr=7, \widetilde{Ra} = 20$. The profiles line up well within the bulk, and reveal a substantial advective heat transport all the way to the edge of the Ekman layer. The steepest $Nu(\widetilde{Ra})$ scaling is observed in the CTC regime with $Pr=7$, where the Ekman flow at both boundaries acts to efficiently increase the heat transport through these vertically coherent structures. \textcolor{black}{For $Pr=1$, where the CTCs are replaced by plumes that mix into the interior, Ekman pumping has a much smaller effect on the heat transfer.}
\begin{figure}
\centering
\includegraphics[width=0.98\linewidth]{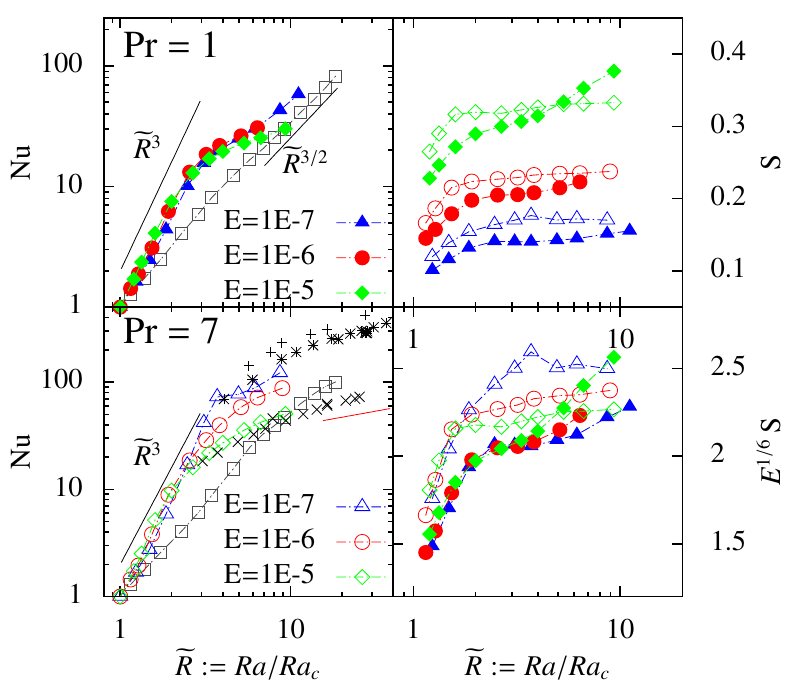}
\caption{Ekman number dependence of $Nu$ and pumping efficiency. Colored symbols represent no-slip DNS results (full symbols: $Pr=1$, open symbols: $Pr=7$), grey symbols show the asymptotic results. The panel in the lower left also contains experimental data ($+$: $2.56\times 10^{-8} \le E \le 2.77 \times 10^{-8}$, $*$:  $0.93 \times 10^{-7} \le E \le 1.15 \times 10^{-7}$, $\times$: $0.8 \times 10^{-5} \le E \le 1.5\times 10^{-5}$ with $5 \le Pr \le 11$ in all cases.).
The right panels show the RMS vertical velocity close to the edge of the Ekman layer, measured at $z_E=3/4 \pi \sqrt{2E}$,  normalized by the RMS vertical velocity at midlayer, a ratio we call $S$ here. Note that at $z=z_E$, classical Ekman layer theory predicts the pumping velocity to be identical to its far-field value, such that values at this location provide a good measure of Ekman pumping.}
\label{fig:figure5}
\end{figure}

For clarity, the presentation so far has focussed exclusively on cases with $E=10^{-7}$. Figure \ref{fig:figure5} illustrates the system behavior over the range $10^{-5} \le E \le 2.5 \times 10^{-8}$ found in no-slip DNS and laboratory experiments. Instead of using $\widetilde{Ra}=Ra E^{3/4}$, here we normalize $Ra$ by the critical value for the onset of convection $Ra_c$ as determined from Chandrasekhars "first approximation" \cite{chandrasekhar1961hydrodynamic,dawes2013comment}, because the asymptotic scaling law $Ra_c  = 8.6956 E^{-4/3}$ becomes inaccurate at larger $E$. 

Surprisingly, in the presence of no-slip boundaries, no obvious convergence towards the asymptotic predictions is observed with decreasing Ekman number. For $Pr=7$, the steep scaling regime extends to higher $\widetilde{R}$ at lower $E$, such that the maximum deviations from the asymptotic theory in fact {\it increase} with decreasing $E$. This trend is observed in both DNS and experiments and likely occurs because vertically coherent structures remain prevalent over a larger super-criticality range at lower $E$. For $Pr=1$, the effect is less pronounced, which is in accord with the fact that plume structures develop at much smaller $\widetilde{R}$ and mix into the interior before reaching the opposite boundary, thus diminishing the effects of pumping. A close inspection of the numerical data shown in figure \ref{fig:figure5} suggests that there might be a weak trend towards  the asymptotic modeling results  for small $\widetilde{R} \lesssim 2$ at both $Pr=1$ and $Pr=7$, but data extending to much lower $E$, at present a significant challenge in the DNS and laboratory setting, is necessary to substantiate this effect.

The panels on the right of figure \ref{fig:figure5} show the relative strength of the Ekman flow in the simulations, measured by a quantity $S$ that we define as the rms vertical velocity at  $z=3/4 \pi \sqrt{2E}$ (i.e. close to the edge of the Ekman layer) normalized by its value at midlayer. The Ekman flow reaches amplitudes roughly between ten and forty percent of the vertical velocity in the bulk, and thus clearly is not negligible. The quantity $S$ increases with $\widetilde{R}$ and decreases with $E$, for moderate $\widetilde{R}$ roughly proportional to $E^{1/6}$, as shown in the lower right panel of figure \ref{fig:figure5}, where the data clusters in two groups depending on $Pr$. The $E^{1/6}$ scaling is expected from (\ref{eq:BCs_pumping}) if we employ the usual estimates $[L]=O(E^{1/3}H)$ and $[u] = O(E^{-1/3} \nu/H)$ for the typical horizontal length scale and flow velocities \cite{julien2012statistical}.


Our results show that in the studied parameter range, the Nu(Ra) scaling is not converging towards the asymptotic modeling results {\it even though} the Ekman flow substantially decreases with decreasing $E$. \textcolor{black}{This finding, and the enormous magnitude by which pumping increases the heat transfer need to be explained by future theory. A positive feedback between thermal boundary layer instabilities and the secondary Ekman flow supporting them may be considered in this context. 
It clearly remains to be seen if} pumping ultimately looses its significance in the very low, but finite Rossby number regime, as presently presumed by theory. \textcolor{black}{Currently, neither experiments nor simulations support this claim.} A promising avenue towards \textcolor{black}{ gaining a better understanding} is the inclusion of parameterized Ekman pumping into the asymptotic theory, as this approach works in our full simulations. Equally important are experiments and DNS reaching still lower $E$ values. Even if pumping effects \textcolor{black}{ should finally become} negligible, figure \ref{fig:figure5} suggests that values several orders of magnitude smaller than $E=10^{-8}$ are needed in order to observe this.  

Another finding of this study is that, similar to the situation in non-rotating convection, heat transport in the low Rossby number regime cannot be characterized by a single, universal scaling exponent $\alpha$. Instead, different dynamical regimes exist, all exhibiting their own heat transfer characteristics, as also suggested by asymptotic theory \cite{julien2012statistical, julien2012heat}. The steep $\alpha \sim 3$ scaling for example breaks down well within the low Rossby number regime ($Ro_c \approx 0.01$ for $Pr=7$, $E=10^{-7}$) when CTCs evolve into plumes. Beyond this transition, a much slower growth of $Nu$ with $\widetilde{R}$ is observed. Unfortunately, laboratory experiments have yet to reach small enough $\widetilde{R}$ at low $E$ to investigate this regime change \cite{king2012heat,ecke2014heat,cheng2014laboratory}. The experimental data presented here are no exception (cf fig. \ref{fig:figure5}). 

The usual motivation for studying rotating convection is its ubiquitous occurrence in geo- and astrophysical systems. Our findings suggest that Ekman pumping, which is typically regarded as negligible in this context \cite{kuang1997earth,roberts2012theory}, may play a more prominent role than previously thought.

\begin{acknowledgments}
S.S. and M.L. conducted direct numerical simulation, K.J. and G.V. contributed asymptotic modeling results and E.M.K., J.C., A.R. and J.A. provided experimental data. We thank the John von Neuman Institute in J\"ulich / Germany for the allotted computer time. S.S. acknowledges funding from the German science foundation under the grant STE1976/1-2. K.J. and J.M.A. were funded by the CSEDI grant \#1067944. \textcolor{black}{Geoff Vasil acknowledges funding from the Australian Research Council, project number DE140101960.}
\end{acknowledgments}

\bibliography{prl_bib}

\end{document}